\documentclass[%
 reprint,
 amsmath,amssymb,
 aps,
]{revtex4-1}

\usepackage{graphicx}
\usepackage{dcolumn}
\usepackage{bm}

\begin{document}

\preprint{APS/123-QED}

\title{A Simple Apparatus for the Direct Measurement of Magnetic Forces and Magnetic Properties of Materials}

\author{Jan A. Makkinje}
\author{George O. Zimmerman}
\affiliation{Boston University}

\date{\today}

\begin{abstract}

In this paper, we describe a simple apparatus consisting of a scale, capable of a one milligram resolution, and a commonly obtainable magnet to measure magnetic forces. This simple apparatus is capable of measuring magnetic properties of materials in either a research or an instructional laboratory. We illustrate the capability of this apparatus by the measurement of the force of iron samples exerted on the magnet, the force of a paramagnetic sample, that by a current carrying wire, and the force of a high temperature superconductor. 

\end{abstract}

\pacs{Valid PACS appear here}
\maketitle


\section{Introduction}

The measurements of magnetic susceptibility (MS) and magnetic properties of materials have contributed substantially to the understanding of the quantum-mechanical nature of condensed matter. But, for most part, the methods for the quantitative measurement of MS have been complicated, cumbersome and costly. Electronic methods rely on the measurement of the change of mutual inductance with some parameter (temperature or magnetic field), and are frequency dependent due to magnetic relaxation times and electrical conductivity \cite{maxwell}, \cite{white}, \cite{loun}. Static measurements rely on the force a magnetic material experiences in a magnetic field \cite{guoy}, \cite{ernshaw}, \cite{connor}, \cite{lewis}. With the availability of high powered and inexpensive permanent magnets, one can easily construct a device for the measurement of magnetic forces on a magnet atop a scale. For many substances, the forces are of the order of milligrams, and since a ferromagnetic or paramagnetic material would attract the magnet, it would lessen the weight, or force the magnet exerts on the scale. Thus the scale needs to have the capability to detect either decrease or an increase in the weight of the magnet. 

\section{Apparatus}

\begin{figure}[h!]
\centering
\includegraphics[height=1.6in]{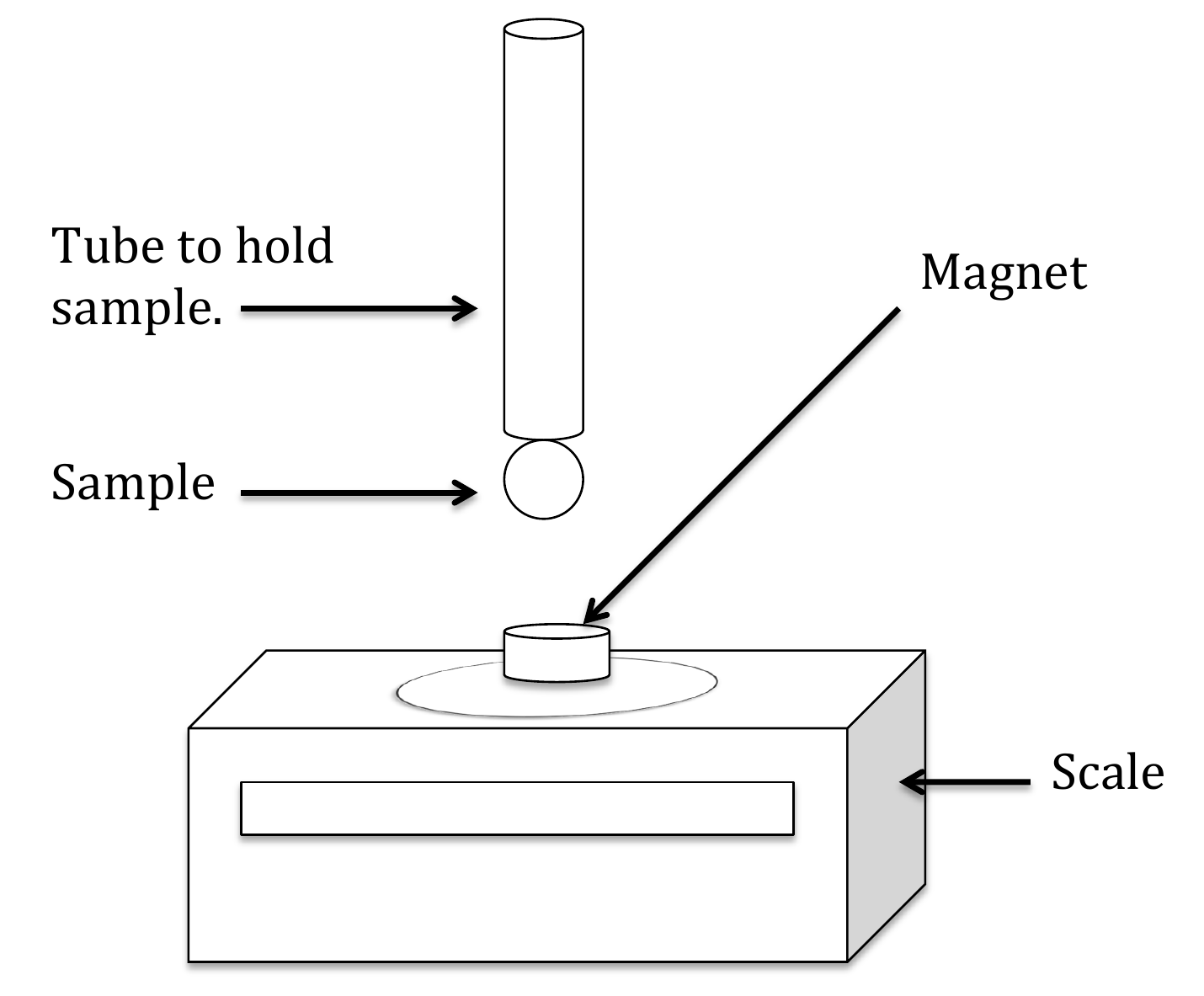}
\caption{The apparatus set up for the measurement of the force of a sample on a magnet on top of a scale with 1mg resolution. The scale, magnet and sample are indicated.}
\label{fig:Apparatus}
\end{figure}

We have constructed an apparatus for the static measurement of magnetic susceptibilities. The apparatus shown in Fig.~\ref{fig:Apparatus} consists of a scale capable of one milligram resolution and a neodymium cylindrical magnet, bought in a toy store, 10mm diameter and 6mm high with a magnetic field of 0.27T at its surface. It is advantageous if the scale comes with a computer interface so that its readings can be recorded together with any other relevant parameters. What one measures is the force on the magnet when a magnetic material is placed at a fiducial point. The force is due to the field which is supplied by the magnet at the location of the sample. Fig.~\ref{fig:Apparatus} shows the general setup with the scale, sample centered above the magnet, and the magnet sitting on the scale. It is prudent to make sure that the pan of the scale is nonmagnetic, since a magnetic pan can distort the field. Since the magnetic field changes drastically with the distance, one also needs a means for the determination of the distance between the sample and the magnet face. In our case this was done by projecting the ray of a laser level on a ruler so we could achieve a one-millimeter accuracy. Another means of measurement could be to rigidly attach the sample to a micrometer drive.

\section{Calculation of the Magnetic Field}

\begin{figure}
\centering
\includegraphics[height=1.6in]{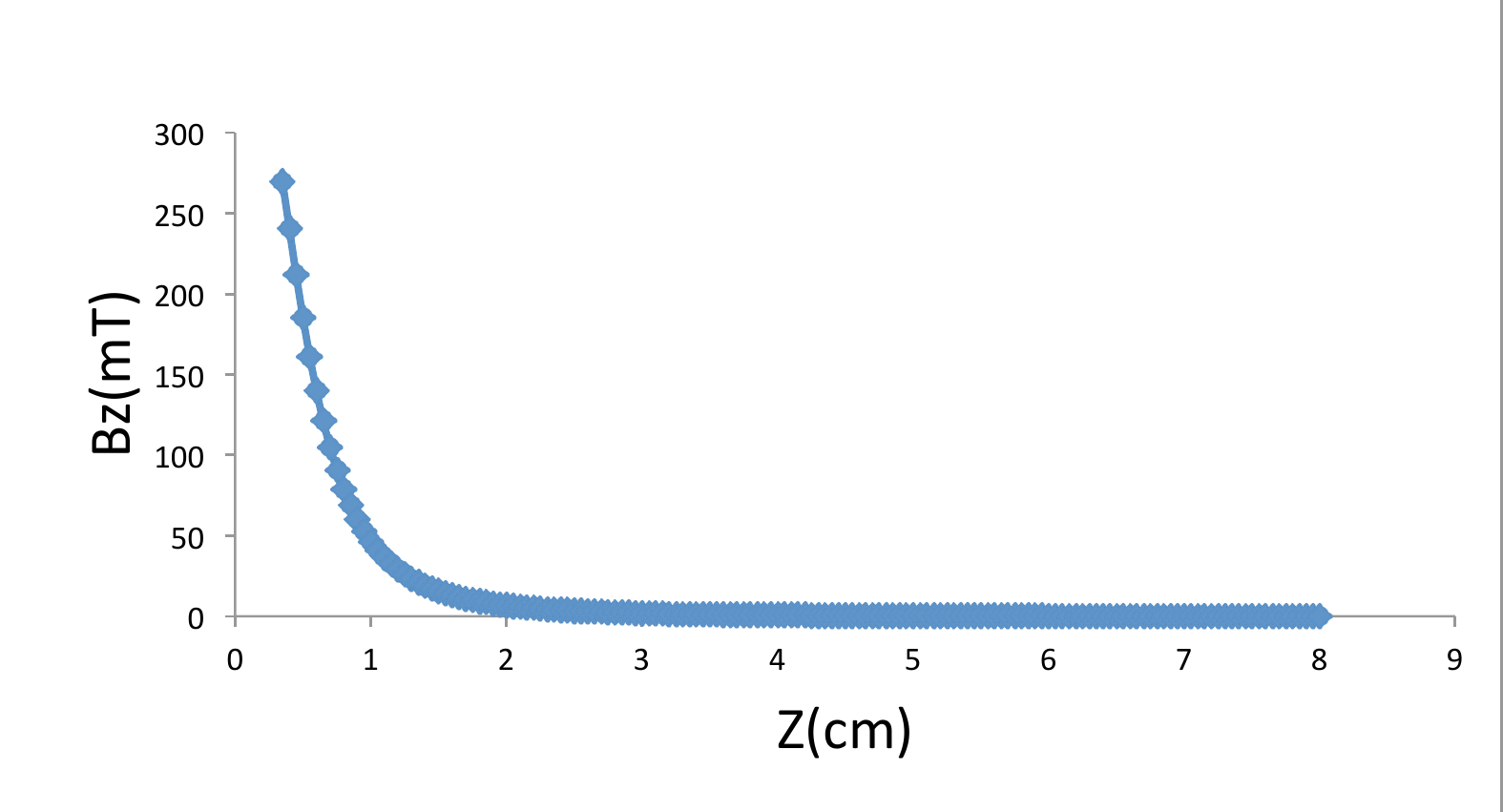}
\caption{The calculated magnetic field of the magnet above the center of the magnet as a function of the distance from the mid plane of the magnet. The origin is assumed at the center of the magnet.}
\label{fig:Fig2}
\end{figure}

\begin{figure}
\centering
\includegraphics[height=1.6in]{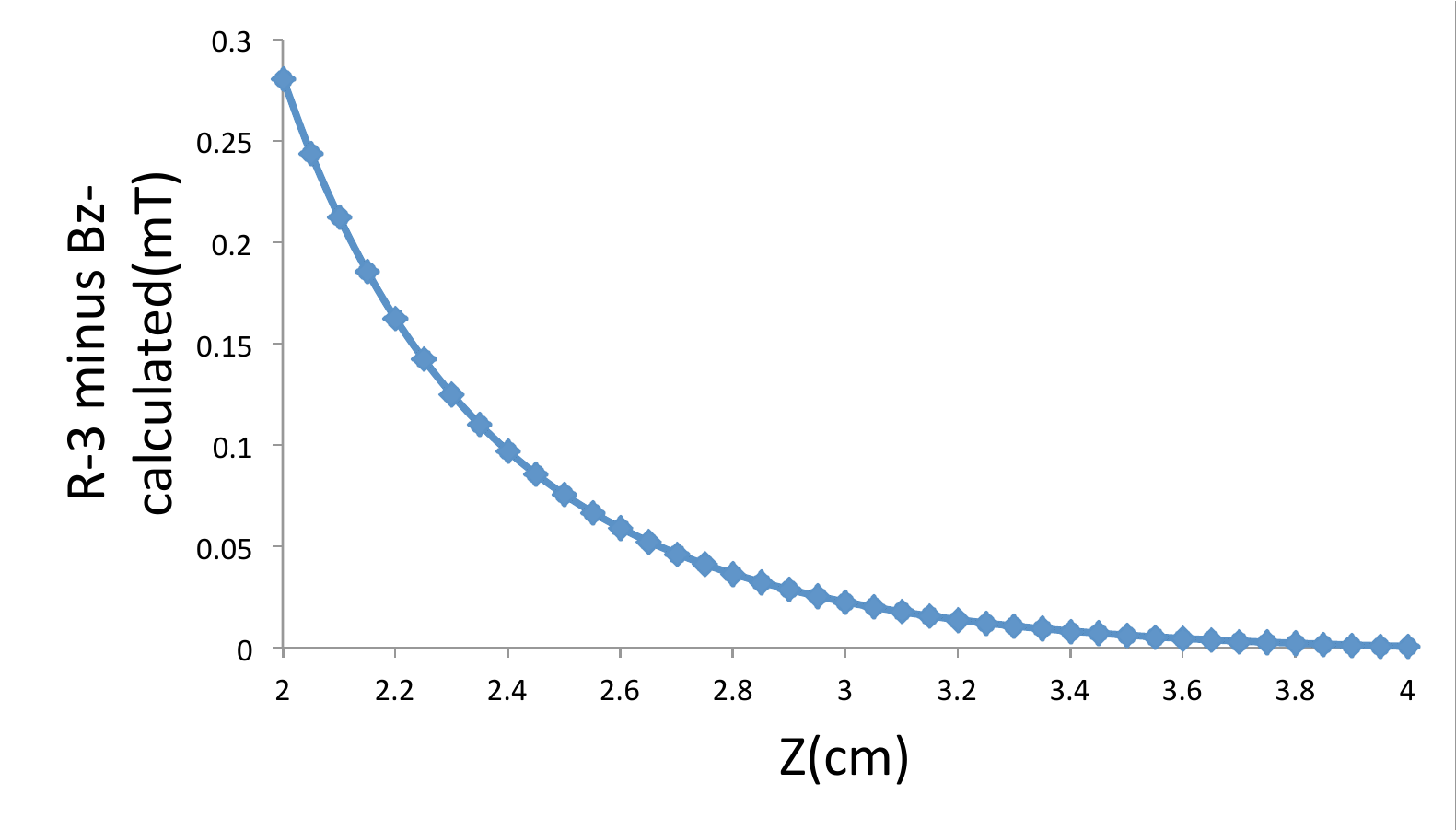}
\caption{Difference between a dipole approximation to the $B_{z}$ field at the center of the magnet and the calculated field. The dipole approximation overestimates the field. See text.}
\label{fig:Fig3}
\end{figure}

The magnetic field was calculated assuming a set of 2000 positive and negative monopoles, separated by 6 millimeters, arrayed on opposite faces of the magnet in a rectangular configuration within a circle of 1 cm diameter. Because the diameter of the magnet is of the same order of magnitude as the distance between the magnet and the sample, a dipolar approximation is not sufficient. Fig.~\ref{fig:Fig2} shows the calculated field $B_{z}$, at the center of the magnet as a function of $Z$, the distance from the $X-Y$ plane. The origin is assumed to be at the center of the magnet. The maximum distance in this calculation is 8cm. That distance was chosen because the magnet contribution of the B-field at that distance is equivalent to that of Earth's contribution. Fig.~\ref{fig:Fig3} shows the deviation of the calculated field from the $\frac{1}{r^{3}}$ dipolar approximation law \cite{Serway} with $r$, equal to $Z$ at $X=Y=0$, being the distance from the magnet. The constant was determined from the best fit at $r$ greater than 3cm. One can see that at a distance smaller than 2cm, the $\frac{1}{r^{3}}$ law deviates significantly from that which is calculated. 

\section{Iron Sample Measurement}

\begin{figure}[h!]
\centering
\includegraphics[height=1.6in]{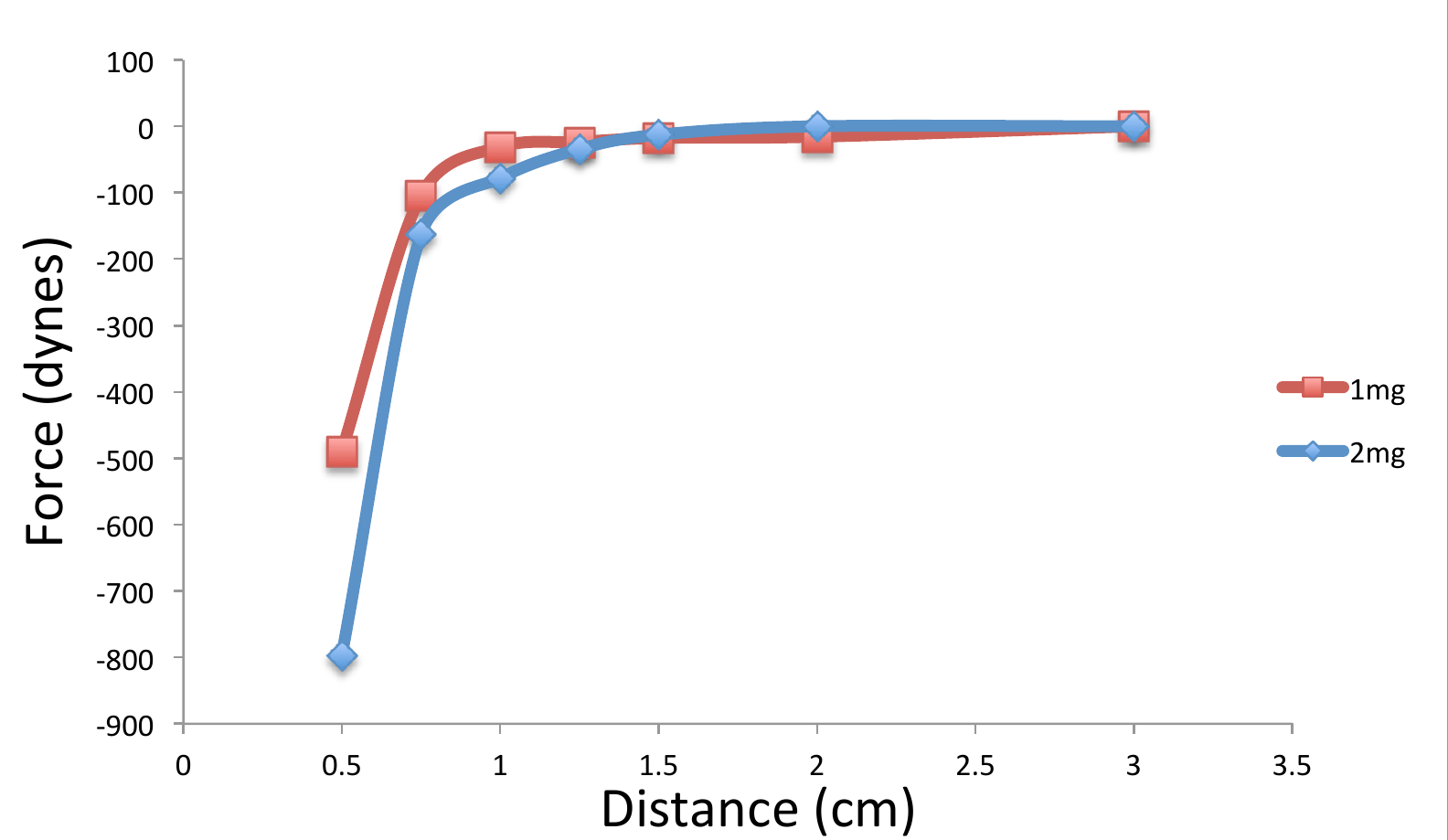}
\caption{Force on the magnet, attraction, of 1mg and 2mg iron samples. Note that the force of the 2 mg sample is about twice that of the 1mg sample. Of course, iron is ferromagnetic.}
\label{fig:Fig4}
\end{figure}

Pure chemical reagent type iron wire was used as the as the sample. Fig.~\ref{fig:Fig4} shows the result of two samples weighing 1mg and 2mg respectively. Since the wire is ferromagnetic, it will attract the magnet, thus making it to appear lighter than if the wire were absent. The figure shows that the negative change in the weight of the magnet, as a function of the sample distance from the magnet, is about twice as large for the 2mg sample as it is for the 1mg sample. The deviation from the ratio of 2 is due to the accuracy with which the samples were weighed since the resolution of the scale was only 1mg. In each case, the force seems to reproduce the magnitude of the magnetic field as a function of distance from the center of the magnet. (A mirror image of the field reflected in the $Z=0$ plane.)

\begin{figure}[h!]
\centering
\includegraphics[height=1.6in]{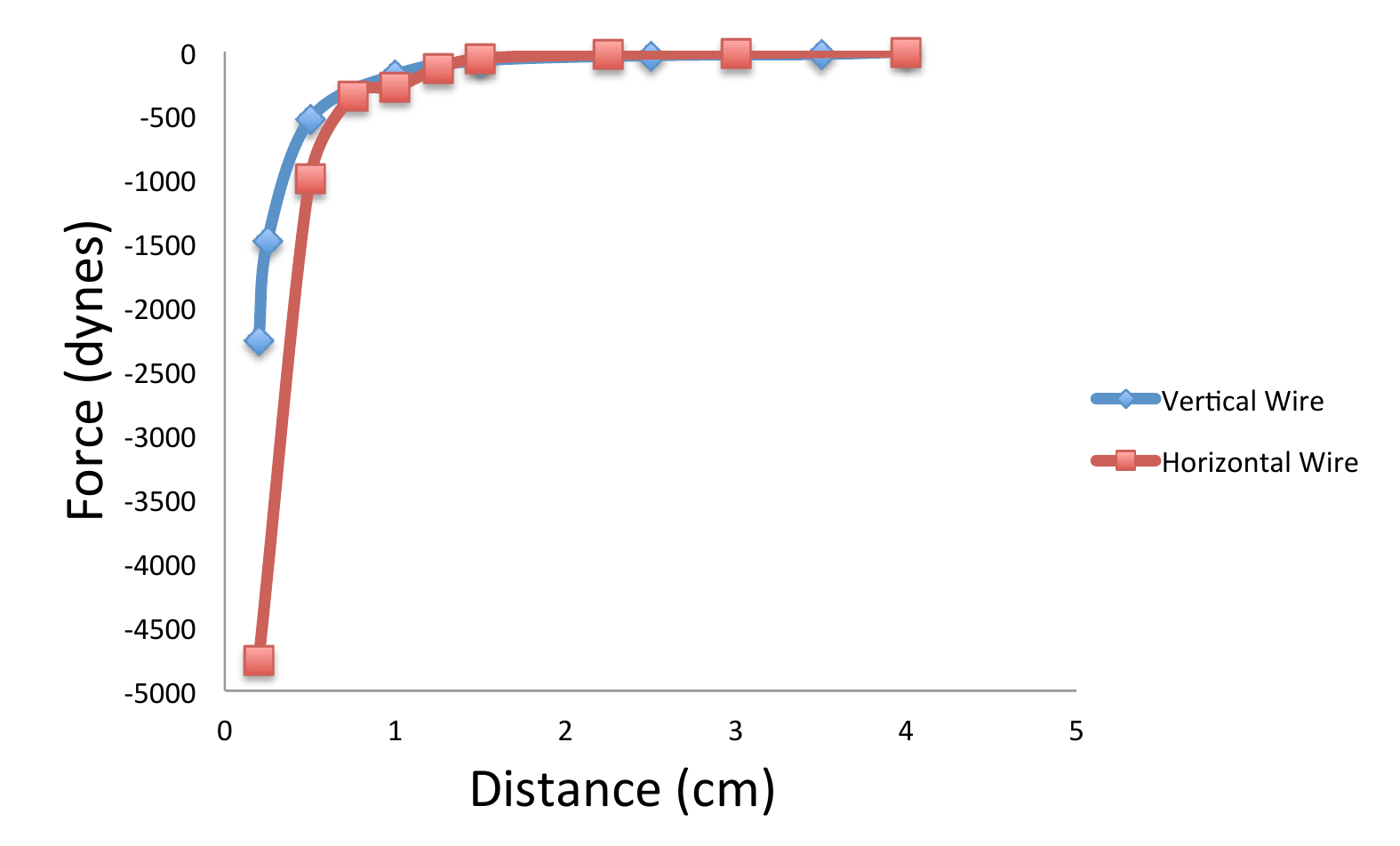}
\caption{Force of an iron wire with its axis parallel, and perpendicular to the magnetic field, as a function of the perpendicular distance. The perpendicular configuration experiences a greater force partly due to the demagnetization factor. The force is attractive resulting in a magnet weight decrease.}
\label{fig:Fig5}
\end{figure}

Next, a 3cm length of iron wire, 0.22mm in diameter was placed vertically in the field and the force measured as a function of the distance of its lowest point from the upper face of the magnet. That measurement was repeated with the wire in the horizontal position. Here we measure the difference of the force due to the demagnetization factor \cite{white}, \cite{loun}, \cite{osborn}. For a long cylinder with the field parallel to the cylindrical axis, the demagnetization factor is zero. If the field is perpendicular to the axis, the demagnetization factor is $\frac{1}{2}$. Fig.~\ref{fig:Fig5} shows the results. The force is smaller with the axis of the wire parallel to the field. The proportions seem of the right order of magnitude, with the difference likely due to the averaging of the $B$ field when the wire is in the horizontal position.

\section{Paramagnetism}

\begin{figure}[h!]
\centering
\includegraphics[height=1.6in]{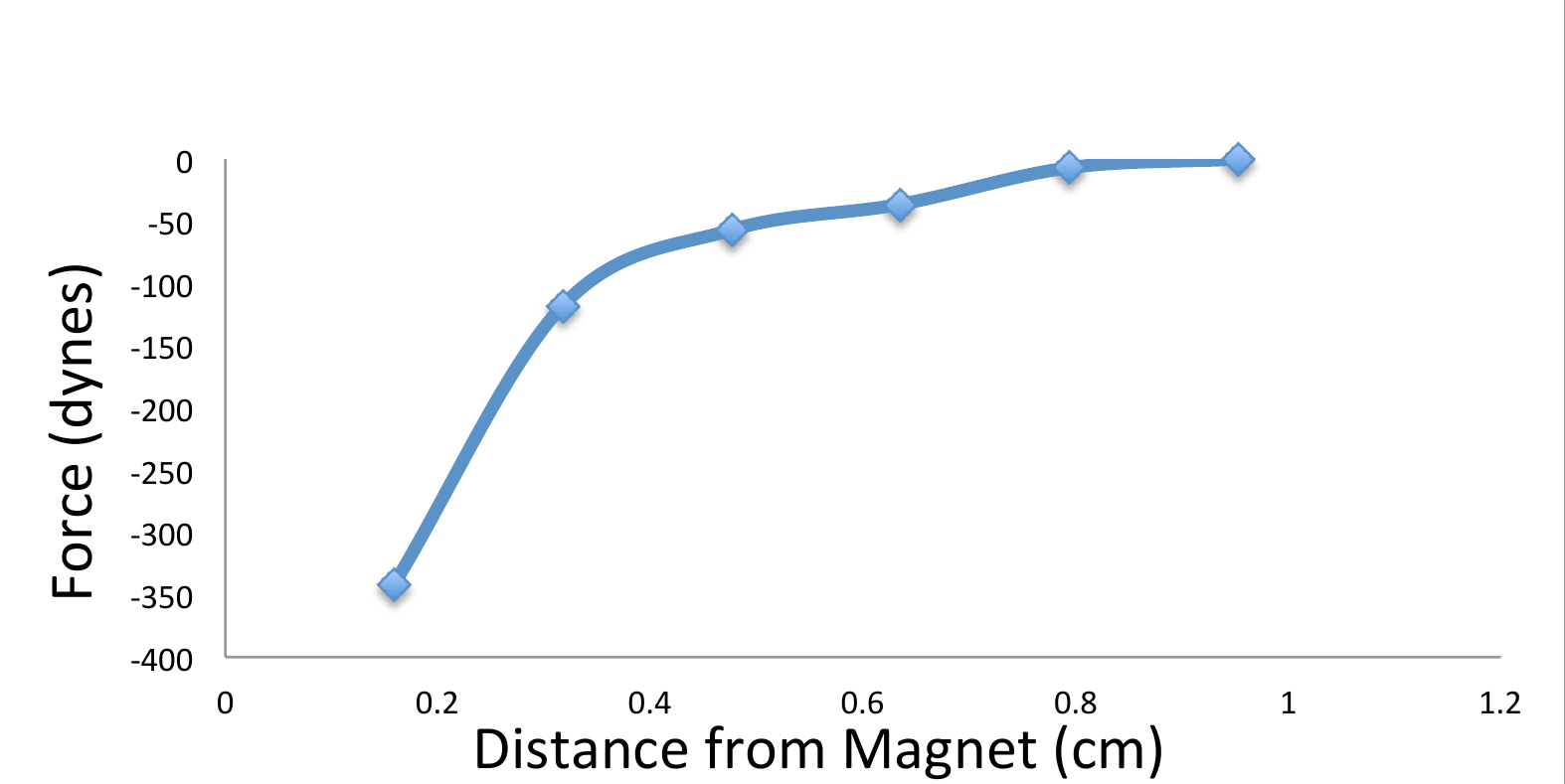}
\caption{Measured force on the magnet by a cylindrical paramagnetic sample of $Gd_{2}O_{3}$ equivalent to a one cgs unit of magnetic susceptibility. See text.}
\label{fig:Fig6}
\end{figure}

A cylindrical sample of $Gd_{2}O_{3}$, 3mm in diameter and 15 mm long, equivalent to one cgs unit of susceptibility was suspended from a string, and its force on the magnet measured as a function of its lower end from the magnet. Again the force shows attraction as seen in Fig.~\ref{fig:Fig6}. It also mimics the calculated field shape. 

\begin{figure}[h!]
\centering
\includegraphics[height=1.6in]{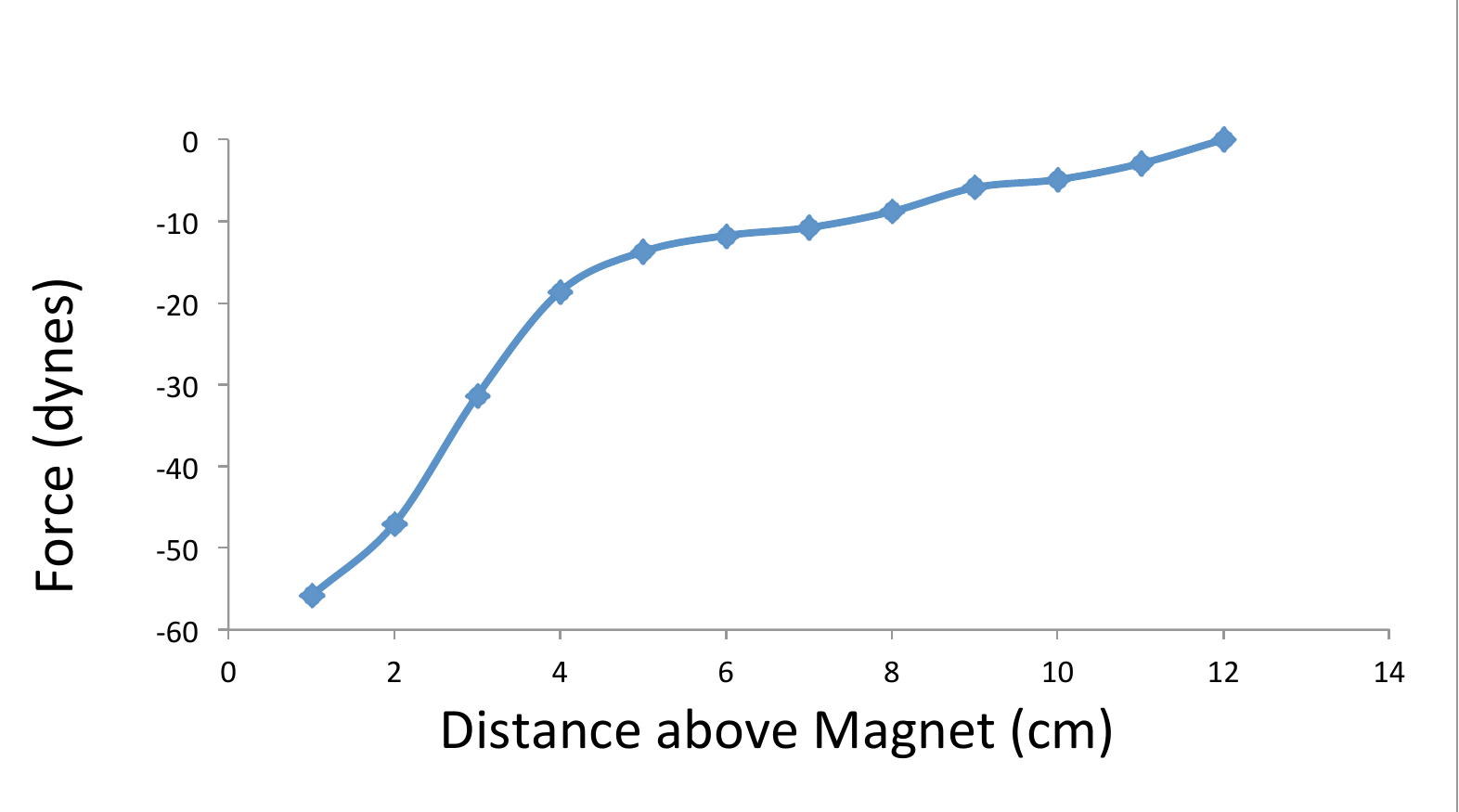}
\caption{Force of a commercial CD ÒJewell CaseÓ. See text.}
\label{fig:Fig7}
\end{figure}

It also turns out that many of the items made out of plastic seem to be paramagnetic. As a sample we measured a plastic commercial CD case, Òjewel caseÓ. That result is shown in 
Fig.~\ref{fig:Fig7}, again as a function of distance from the magnet. Here the force seems to vary somewhat slower than in the case of the iron samples. The surface of the jewel case was in the $X-Y$ plane and was averaged over a much larger area.

\section{Current-Carrying Wire}

\begin{figure}[h!]
\centering
\includegraphics[height=1.6in]{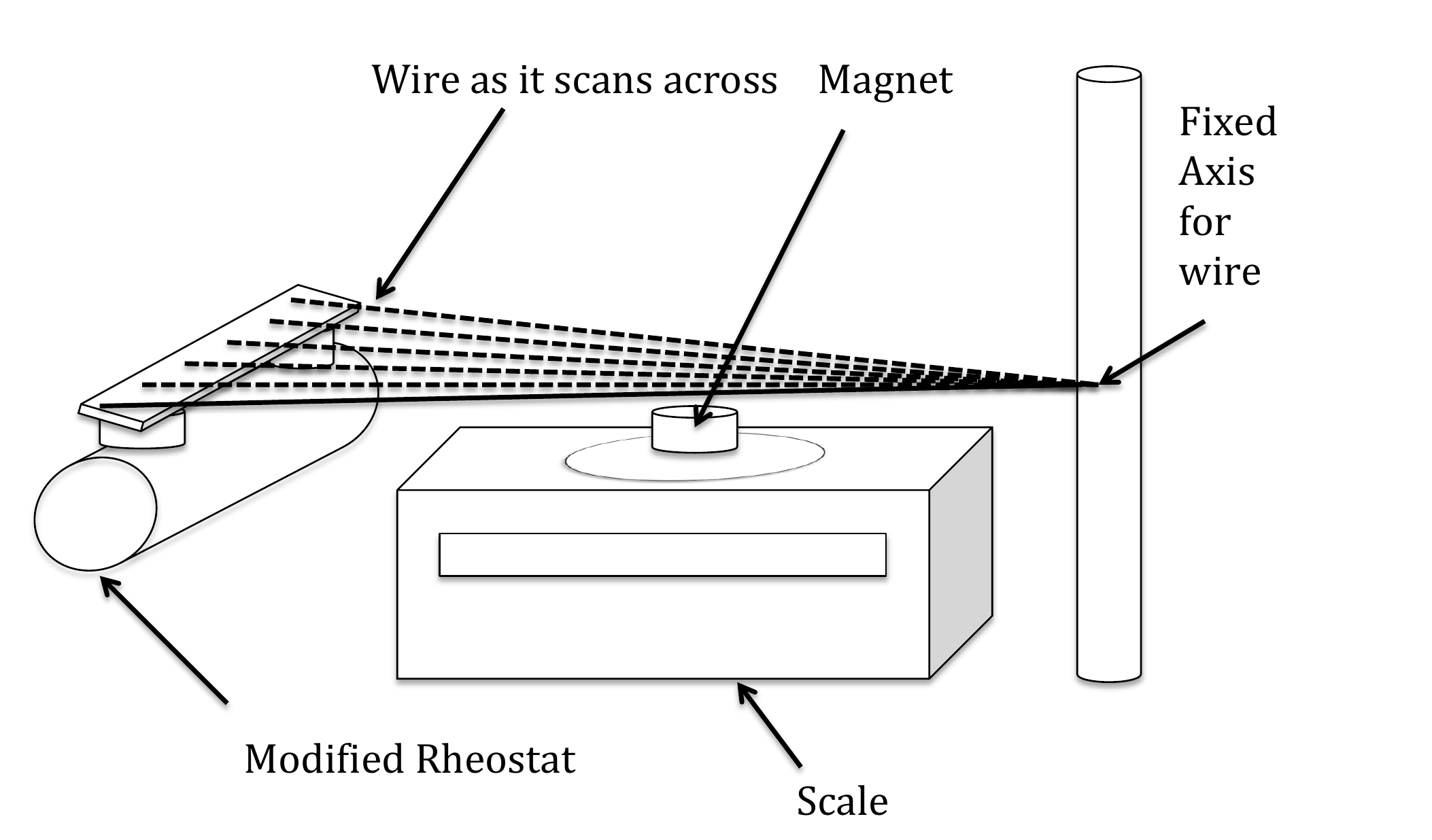}
\caption{Apparatus for the measurement of the force of a current carrying wire, dashed lines, on the magnet. Shown are the scale, the magnet, the wire and the rheostat which, with the wire attached to the slider, gives a voltage proportional to the horizontal position of the wire.}
\label{fig:Fig8}
\end{figure}

Fig.~\ref{fig:Fig8} shows the set up for the measurement of the force exerted on the magnet by a current carrying wire. The scale with the magnet sitting on top of it, are indicated. As previously described, the magnet is again 6mm high and 10mm in diameter with a magnetic field of 0.27T as measured on its face. The dashed lines denote the current carrying wire which is moved horizontally and which exerts a force on the magnet. One end of the wire is fixed while the other is mechanically tied to the slider of a rheostat. The rheostat supplies a voltage which depends on the position of the slider when a battery is placed across the rheostat. In that way, a signal proportional to the position of the wire, tied to the slider, can be recorded at the same time as the force on the scale.

\section{Lorentz Force}

As can be found in any introductory physics book, (\cite{Serway}, eq.(31.16)), the force exerted by a magnetic field on a current-carrying wire in the absence of electrostatic charges is

\begin{equation}
\vec{F} = l\vec{I} \times \vec{B}
\end{equation}

where $F$ denotes the force and $I$ and $B$ are the current and the magnetic field, respectively. We assume that the $+Z$ direction is along the magnetic moment at the center of the magnet, and the current is along the $Y$-axis. The force on the magnet is measured as the wire changes its position along the $X$-axis. Actually, there is a slight angle between the $Y$ direction and the wire, but because of the geometry, the deviation from $Y$ is minimal. As previously stated, one end of the wire is mechanically attached to a vertical rod, while the other is mechanically attached to the slider of a rheostat. An upward force will give a negative weight indication while downward force gives a positive change. Because we are only measuring the vertical force $F_{z}$ and assume the current to be $I_{y}$, only the $-I_{y}B_{x}$ component is measured. The force was then summed along the length of the wire. 

\begin{figure}[h!]
\centering
\includegraphics[height=1.6in]{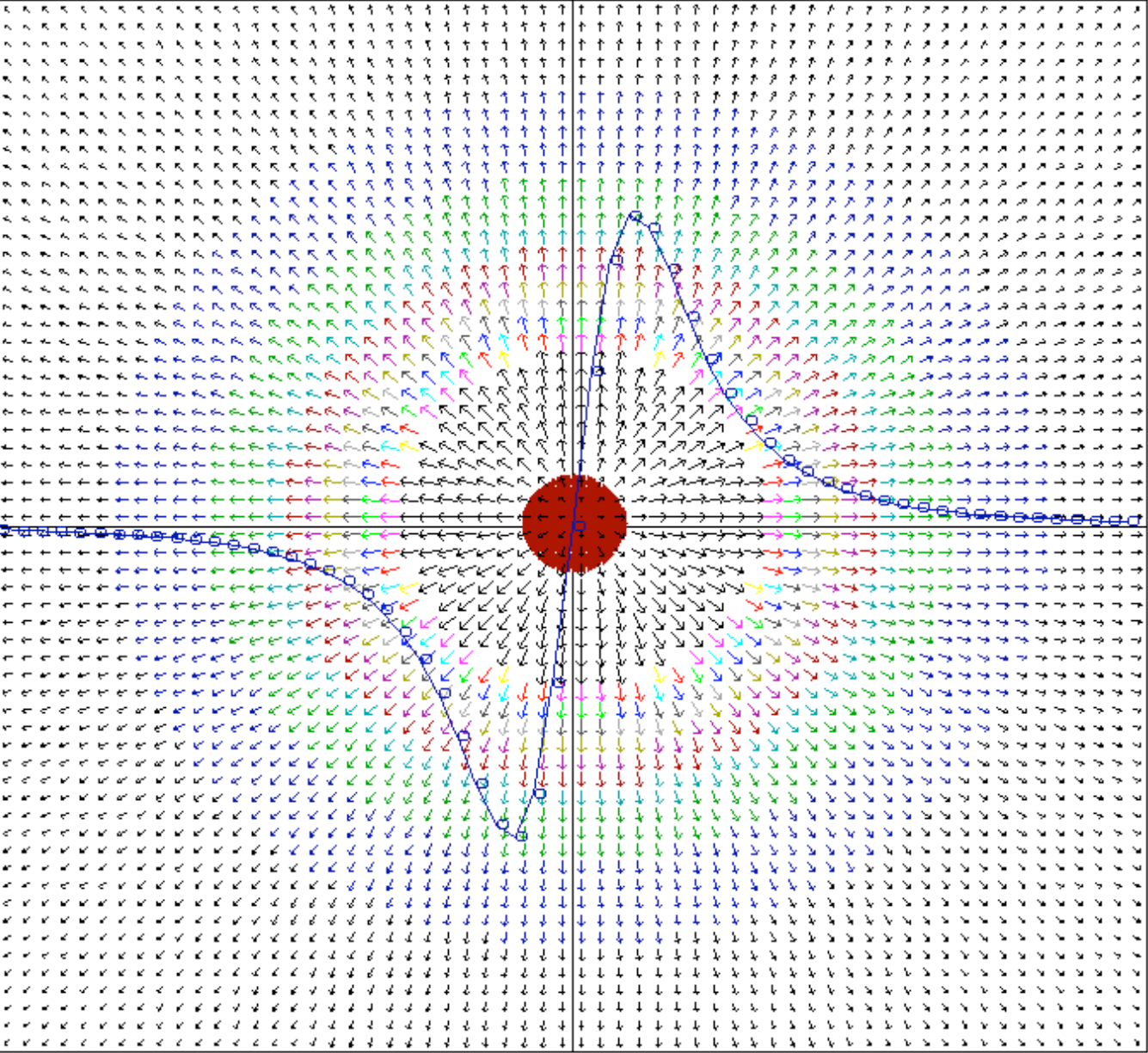}
\caption{Calculated $X-Y$ components of the magnetic field as one looks down on the $X-Y$ plane. The circle denotes the magnet, while the curve denotes the force of the wire on the magnet. }
\label{fig:Fig9}
\end{figure}

As before, the magnetic field was calculated assuming a set of 2000 positive and negative monopoles, arrayed on opposite faces of the magnet in a rectangular configuration within a circle of 1 cm diameter. Fig.~\ref{fig:Fig9} shows the $X$ and $Y$ components of the field as seen by looking down on the magnet. The positive, north-pole, is pointing up along the $+Z$ direction, while the viewer is looking down, along the $-Z$ axis. The length of the arrows indicates the $X-Y$ plane intensity of the $B$ field, and the arbitrary colors give a concept of the total field intensity. The superimposed curve and blue points denote the $Z$-component of the force due to the current on the magnet. The current carrying wire is parallel to the $Y$-axis, up.

\begin{figure}[h!]
\centering
\includegraphics[height=1.6in]{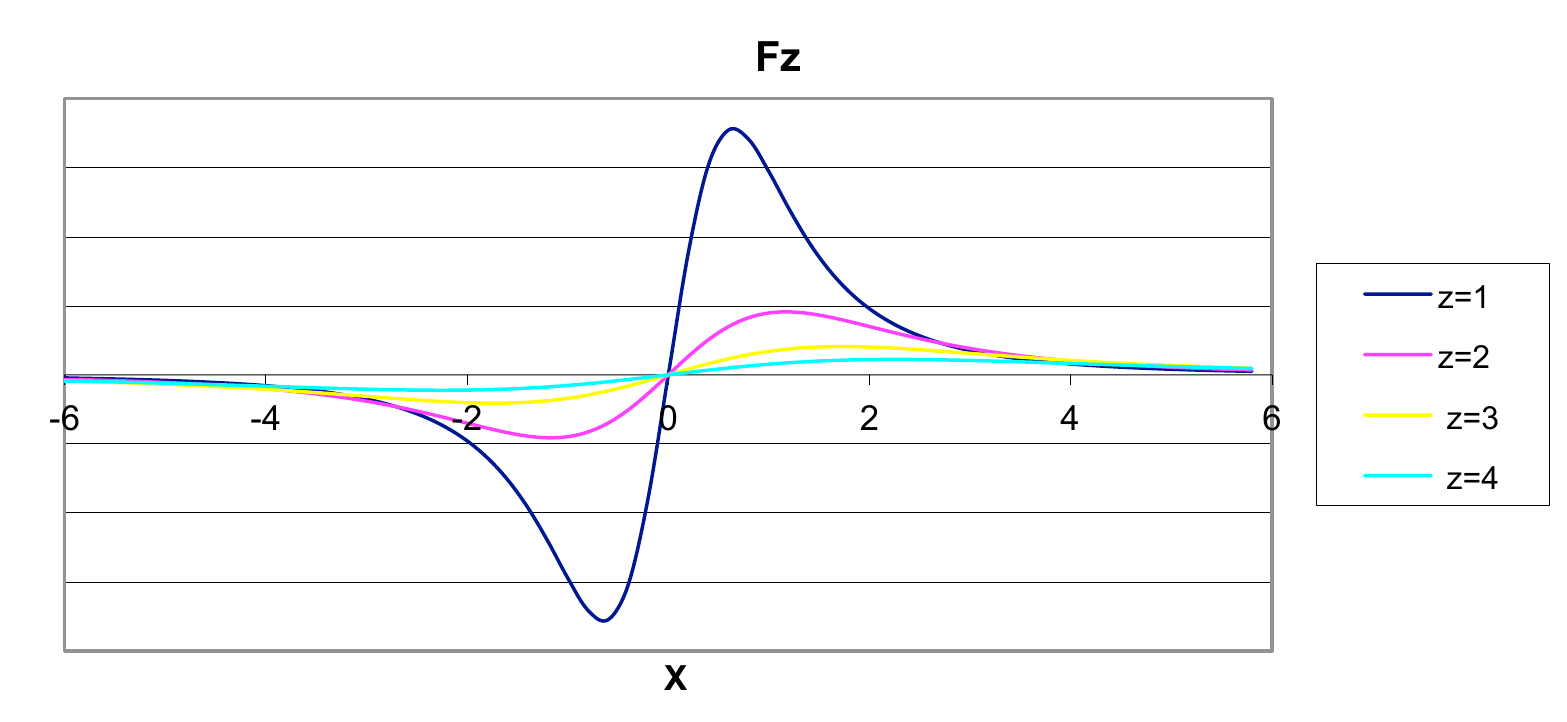}
\caption{Force of the wire on the magnet as a function of $X$, the distance of the wire from the center of the magnet, for several elevations $Z$ above the magnet. The current is in the $Y$ direction, up.}
\label{fig:Fig10}
\end{figure}

Fig.~\ref{fig:Fig10} gives the force due to a current carrying wire for several elevations $Z$, of the wire, as a function of $X$, the horizontal distance from the center of the magnet at a constant elevation, $Z$. Note that the force maxima and minima shift with the $Z$ distance. The units are arbitrary.

\section{Measurement of Current Carrying Wire}

\begin{figure}[h!]
\centering
\includegraphics[height=1.6in]{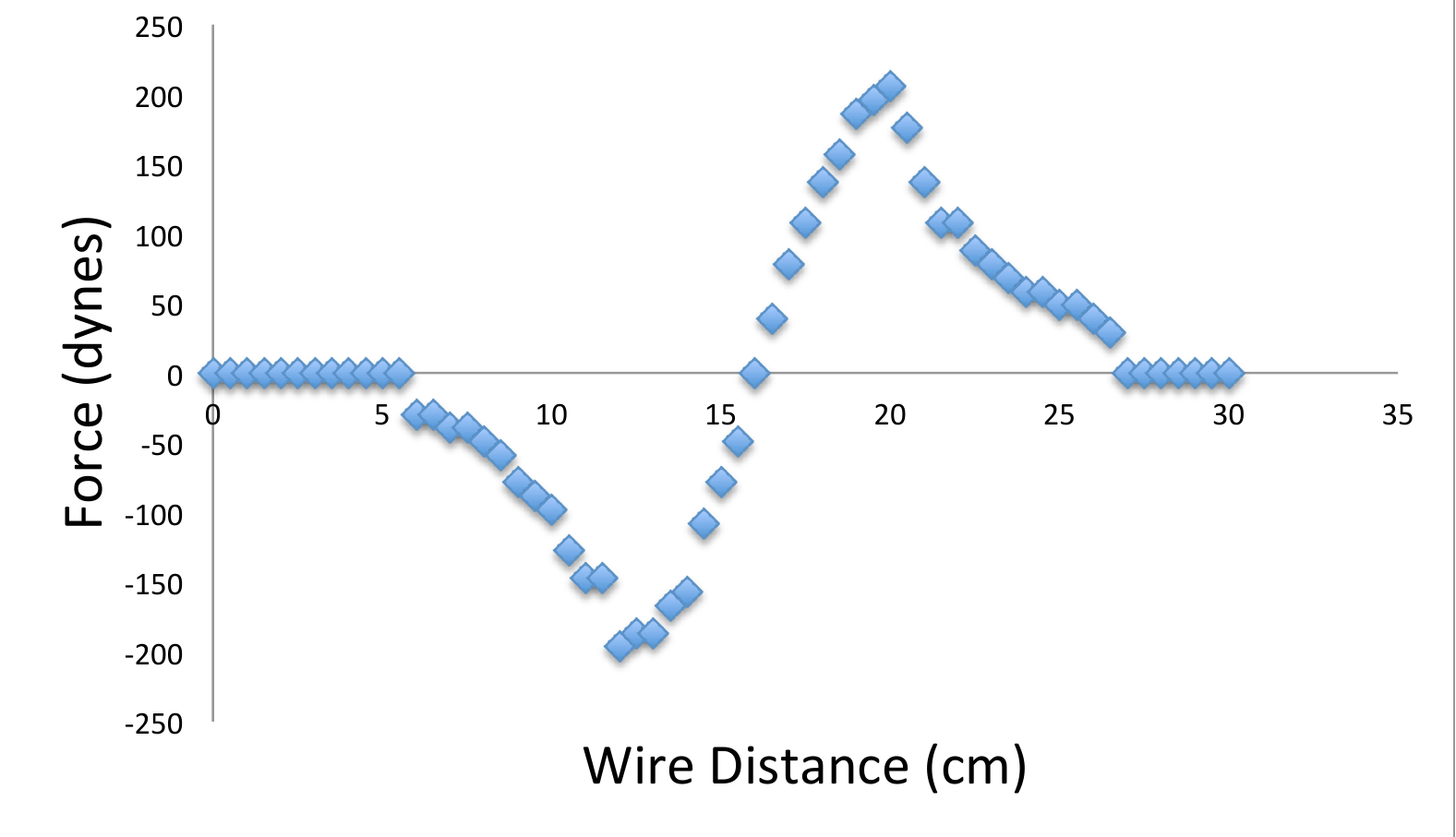}
\caption{Measurement of the force of a current carrying wire on the magnet at constant $Z$ = 2cm, as a function of $X$. $Y$ is the direction of the current. One can see the essential features of the calculations reproduced in the measurements. See text.}
\label{fig:Fig11}
\end{figure}

Fig.~\ref{fig:Fig11} shows the measurement of the force of a wire, with a 4 ampere current, as a function of $X$, the distance from the center of the magnet, at an elevation of 2cm. One can easily see the features shown in the calculations: the zero force far away; an attraction which has a maximum at about one centimeter; the force going to zero when the wire crosses the center of the magnet; and the reversal of the force and its maximum as $X$ increases.  

\section{Measurement of Superconducting Material Diamagnetism}

A beaker with YBCO-123 material in the form of a 3cm diameter disk of mass 13.766g was placed 1.27cm above the magnet. YBCO-123 is a high transition temperature superconductor \cite{tp}. The sample was cooled to below its superconducting transition temperature, which was 96K, by pouring liquid nitrogen into the beaker and then letting it evaporate. The temperature was measured by means of a thermocouple. Caution should be taken that the placement of the thermocouple be such that their interaction with the magnetic field is minimal, since it itself exhibits magnetic properties as a function of temperature. 

\begin{figure}[h!]
\centering
\includegraphics[height=1.6in]{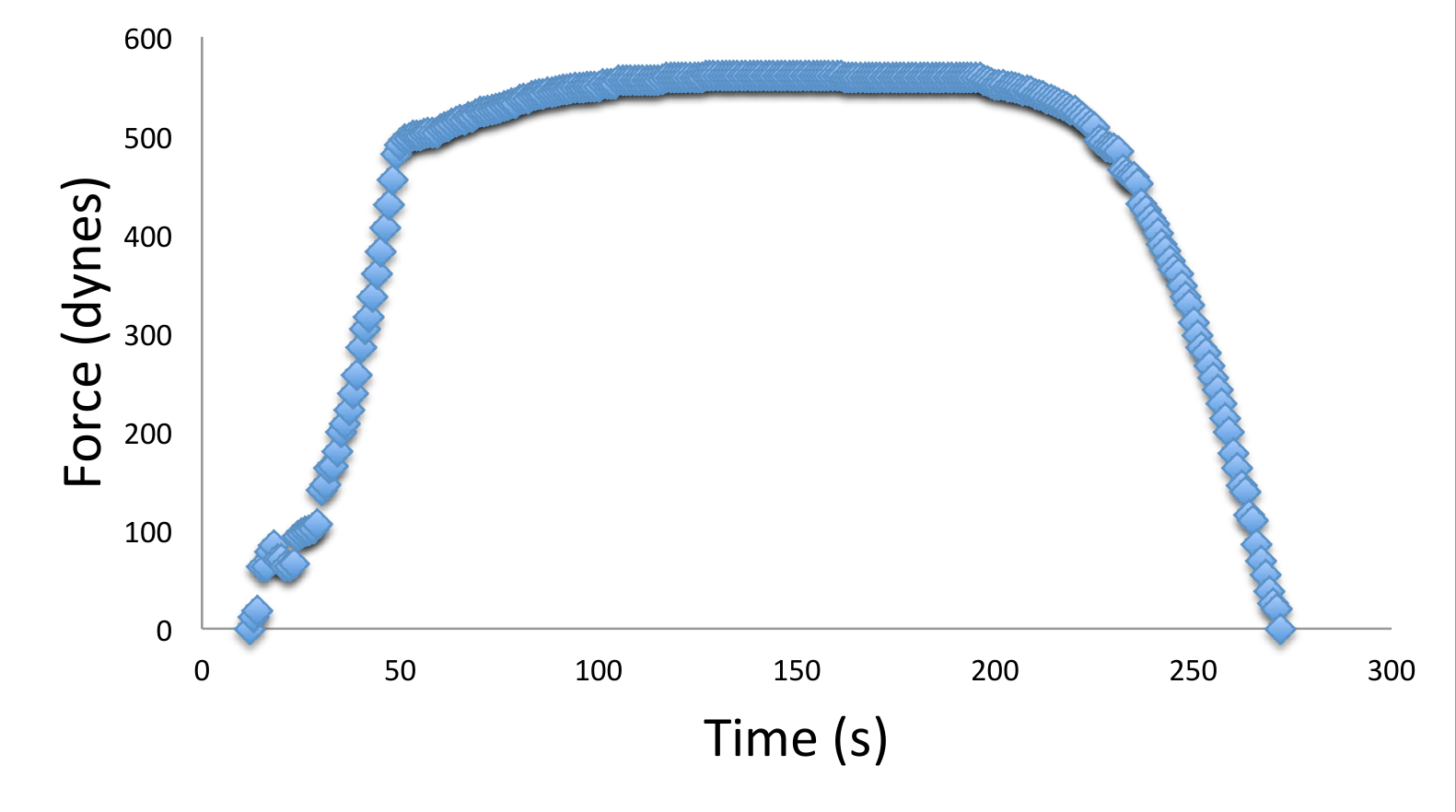}
\caption{The force of a $YB_{2}C_{3}O_{7-d}$ high transition temperature superconductor on the magnet as it cools while liquid nitrogen is poured into the sample beaker. One can see the subsequent warming of the sample while the liquid nitrogen evaporates and the sample loses its superconductivity. The force is repulsive. See text.} 
\label{fig:Fig12}
\end{figure}

Fig.~\ref{fig:Fig12} shows the force exerted by the superconducting material as a function of time. Note, that unlike in the case of the ferromagnetic or paramagnetic materials, the force is positive, showing a repulsion between the material and the magnet. That is due to the fact that a superconductor is highly diamagnetic, exhibiting the Meissner Effect \cite{fritz}. Initially at room temperature, the sample cools to 77K as liquid nitrogen is poured into the beaker. The sample then becomes superconducting and, because of the Meissner Effect, repels the magnet which results in an increase in the weight of the magnet. As the liquid nitrogen evaporates, the sample warms and becomes normal. That results in a steep decrease in the repulsion. Subsequent force is that of the normal phase of YBCO as the sample warms.

\section{Conclusions}

We have demonstrated a versatile, inexpensive apparatus for the measurement of magnetic properties of materials. We have illustrated its versatility by the measurement of the magnetic properties of various samples of ferromagnetic, paramagnetic and diamagnetic samples.

\section{Acknowledgments}

We want to thank Mr. Erich Burton and the Boston University Physics Department for help with aspects of the research reported here.

\end{document}